\documentclass{WileyMSP-template}
\usepackage{soul,xcolor}
\usepackage{amssymb}
\usepackage{hyperref}
\usepackage{setspace}
\begin{document}

\title{Mapping Reversal Pathways and Interaction Fields in Artificial Spin Ice}

\maketitle

\author{Brindaban Ojha $^{1, *}$},
\author{Matías P. Grassi $^{1}$},
\author{Vassilios Kapaklis $^{1, *}$}

\begin{affiliations}
$^{1}$Department of Physics and Astronomy, Uppsala University, Box 516, 75120 Uppsala, Sweden\\
Brindaban Ojha \texttt{https://orcid.org/0000-0002-0276-6937}\\
Matías P. Grassi \texttt{https://orcid.org/0000-0001-9551-9793}\\
Vassilios Kapaklis \texttt{https://orcid.org/0000-0002-6105-1659}
\end{affiliations}
$^{*} $Corresponding author: brindaban.ojha@physics.uu.se; vassilios.kapaklis@physics.uu.se


\begin{abstract}

In artificial spin ice (ASI), magnetic interactions between nanomagnets determine both the stable states and the switching pathways under an applied field. Here, first-order reversal curve (FORC) measurements are used to map how these interactions govern magnetization reversal in square arrays as the element shape and spacing are varied. The FORC diagrams show that some geometries reverse more uniformly, whereas others exhibit broader, more asymmetric responses, indicating stronger interaction effects and more complex reversal pathways. Combined FORC analysis and micromagnetic simulations also capture subtle changes in internal magnetization textures during switching, linking local behavior within individual elements to collective behavior across the array. These results establish FORC as a practical tool for mapping and engineering interaction landscapes, with direct relevance to reconfigurable magnetic reservoirs and neuromorphic functionality.

\end{abstract}


\section{Introduction}

Artificial spin ice (ASI) systems are lithographically defined arrays of nanoscale ferromagnetic elements arranged in well-controlled geometries \cite{skjaervo2020advances}. These engineered structures provide a versatile platform for tuning magnetic interactions and dynamic behavior through deliberate control of lattice geometry and element dimensions \cite{park2017magnetic,gliga2017emergent,stopfel2018magnetic,qi2008direct}. The magnetization state of each nano-island is governed by the interplay between intrinsic shape anisotropy and magnetostatic (dipolar) interactions with neighboring elements. As a result, ASI systems enable the study of a broad range of emergent phenomena, including geometrical frustration \cite{gilbert2014emergent,nisoli2017deliberate,puttock2022defect,RevModPhys.85.1473}, effective magnetic monopole excitations connected by Dirac strings \cite{zhang2021string}, collective dynamics \cite{jungfleisch2016dynamic,lendinez2020magnetization,kapaklis2014thermal}, phase transitions and Coulomb phases \cite{sendetskyi2019continuous,anghinolfi2015thermodynamic,perrin2016extensive,ostman2018interaction, leo2018collective}, which are often more directly observable than in naturally occurring spin-ice materials because individual microstates in ASI can frequently be accessed by imaging techniques. Beyond their fundamental significance, ASI architectures have attracted considerable interest for potential applications in high-density data storage \cite{wang2016rewritable,liu2016nano}, magnetic logic \cite{kaffash2021nanomagnonics,arava2018computational}, nonvolatile memory \cite{caravelli2022artificial,taniguchi2025echo}, and neuromorphic computing \cite{gartside2022reconfigurable}. Recent studies have further highlighted field-controlled operation, fading-memory behavior, and reconfigurable reservoir functionality in ASI-based platforms \cite{gartside2022reconfigurable,jensen2024clocked,penty2025controllable,bhandari2025clocking}.

The magnetic behavior of ASI arrays is highly sensitive to geometric parameters such as nanomagnet dimensions, lattice spacing, and film thickness \cite{strandqvist2025nanomagnet}. These parameters provide effective means of modulating dipolar coupling strength and, consequently, the collective response of the system. Although significant progress has been made in understanding these systems, a comprehensive description of how interaction-field distributions govern magnetization reversal—particularly in regimes dominated by collective effects—remains an open challenge. Experimentally probing such interactions is nontrivial: conventional imaging techniques are often limited in either spatial resolution or field of view, whereas standard hysteresis measurements provide only averaged information.

The first-order reversal curve (FORC) technique is a well-established method for probing magnetization processes in magnetic nanostructures \cite{pike1999characterizing,roberts2014understanding,pike2003first,pohlit2016first}. It provides detailed information on irreversible switching events, magnetic interactions, phase transitions, and distributions of magnetic properties that are often obscured in conventional hysteresis measurements \cite{roberts2014understanding}. Initially introduced in geomagnetism \cite{roberts2014understanding}, the FORC method has since been extended to a wide range of magnetic systems, including nanoparticle ensembles, permanent magnets, patterned nanostructures such as nanodot and antidot arrays \cite{rivelles2025reading}, and thin films \cite{gilbert2021forc}.

In this work, FORC measurements are used to investigate magnetization-reversal mechanisms in square artificial spin ice systems with fixed nanomagnet length while systematically varying width and lattice pitch (periodicity). The analysis examines how these geometric parameters influence both switching fields and interaction-field distributions, giving rise to distinct features in the FORC diagrams. In addition, micromagnetic simulations support the experimental findings and provide deeper insight into how interactions shape magnetic textures, which are directly reflected in the observed FORC response.

\section{Experimental Details}

\begin{figure}[h!]
\centering
\includegraphics[width=0.6\linewidth]{"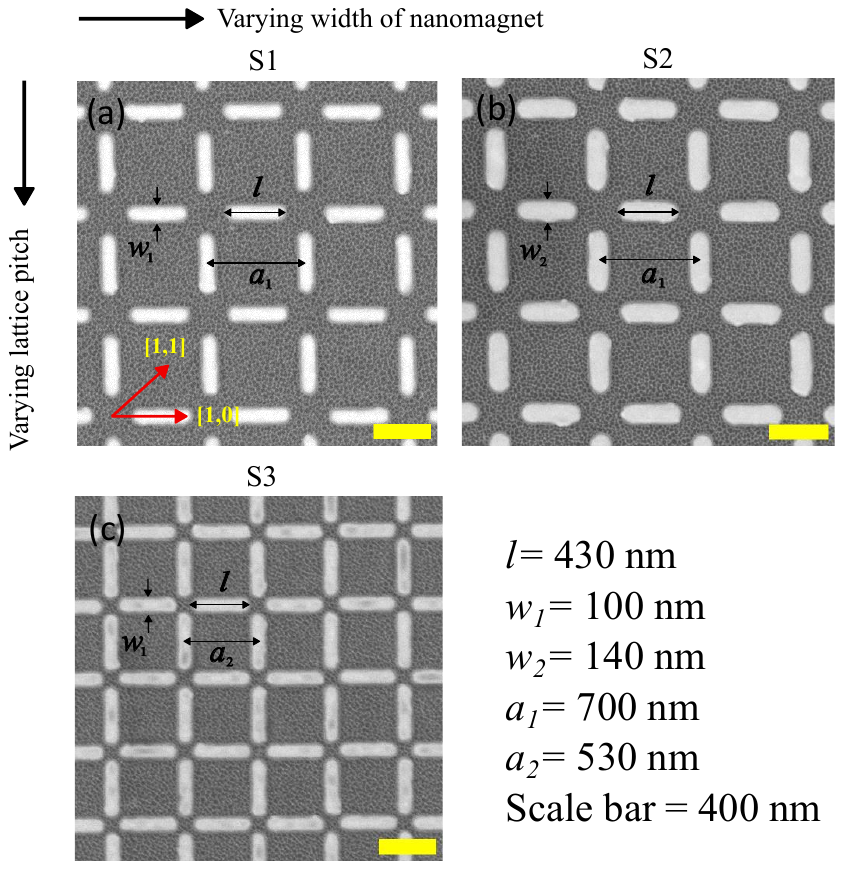"}
\caption{\linespread{1.0} Scanning electron microscopy images of square artificial spin ice arrays patterned by electron-beam lithography (EBL). The nanomagnets have lateral dimensions of $l_1 = 430$ nm, $w_1 = 100$ nm, and $a_1 = 700$ nm for S1 (a); $l_1 = 430$ nm, $w_1 = 140$ nm, and $a_1 = 700$ nm for S2 (b); and $l_1 = 430$ nm, $w_1 = 100$ nm, and $a_1 = 530$ nm for S3 (c). During FORC measurements, the field was applied along the [1,1] direction of the lattice. The scale bar is 400 nm.}
\label{Fig:1}
\end{figure}

A multilayer thin-film stack of Ta (3 nm)/Permalloy (Py) (Ni$_{80}$Fe$_{20}$) (20 nm)/AlO$_x$ (4 nm) was deposited on a Si substrate by magnetron sputtering at a base pressure of $\sim 3 \times 10^{-9}$~mbar. The Ta and Py layers were deposited by DC sputtering, whereas the AlO$_x$ layer was deposited by RF sputtering. The Ta layer served as a seed layer to promote high-quality growth, while the AlO$_x$ capping layer prevented oxidation of the magnetic Py layer. Following deposition, the films were patterned into square artificial spin ice structures by electron-beam lithography and Ar$^{+}$ ion milling. The fabricated nanostructures were characterized by scanning electron microscopy (SEM).

Three types of square artificial spin ice arrays were prepared by varying the nanomagnet width ($w$) and lattice periodicity or pitch ($a$) while keeping the length ($l$) constant, as shown in Figure~\ref{Fig:1}. These arrays are denoted S1, S2, and S3. The three systems were designed to distinguish the effects of intra-island and inter-island interactions on the FORC distributions. Increasing the island width in S2 probes intra-island effects, whereas reducing the lattice pitch in S3 relative to S1 enhances inter-island interactions. The nanomagnets had widths of  100 nm and  140 nm and a common length of 430 nm. The lattice periodicity was set to either 530 nm or 700 nm. Sample S1 had an aspect ratio (AR) $(l/w) \approx 4.3$ with periodicity $a_1 = 700$ nm (Figure~\ref{Fig:1}(a)); S2 had AR $\approx $ 3  with the same periodicity, $a_1 = 700$ nm (Figure~\ref{Fig:1}(b)); and S3 had AR $\approx $ 4.3 with a reduced periodicity, $a_2 = 530$~nm (Figure~\ref{Fig:1}(c)).

Magneto-optical Kerr effect (MOKE) measurements were used to perform first-order reversal curve (FORC) protocols, and the FORC data were processed using the doFORC method \cite{cimpoesu2019doforc}. During these measurements, the field was applied along the [1,1] direction. In addition, micromagnetic simulations were carried out using MuMax3 to elucidate the mechanisms responsible for the experimental observations \cite{vansteenkiste2014design}. In the simulations, an array of 36 nanomagnets was considered, and the cell size was set to $2\times 2\times 20$~nm$^3$, which is below the exchange length. Material parameters corresponding to Permalloy (Py) were used, with saturation magnetization $M_s = 7.5\times10^6$~A/m and exchange stiffness $A = 13$~pJ/m, as reported in the literature \cite{gartside2022reconfigurable}. Periodic boundary conditions were applied in all simulations.

\section{Results and Discussion}

\begin{figure}[h!]
\centering
\includegraphics[width=0.6\linewidth]{"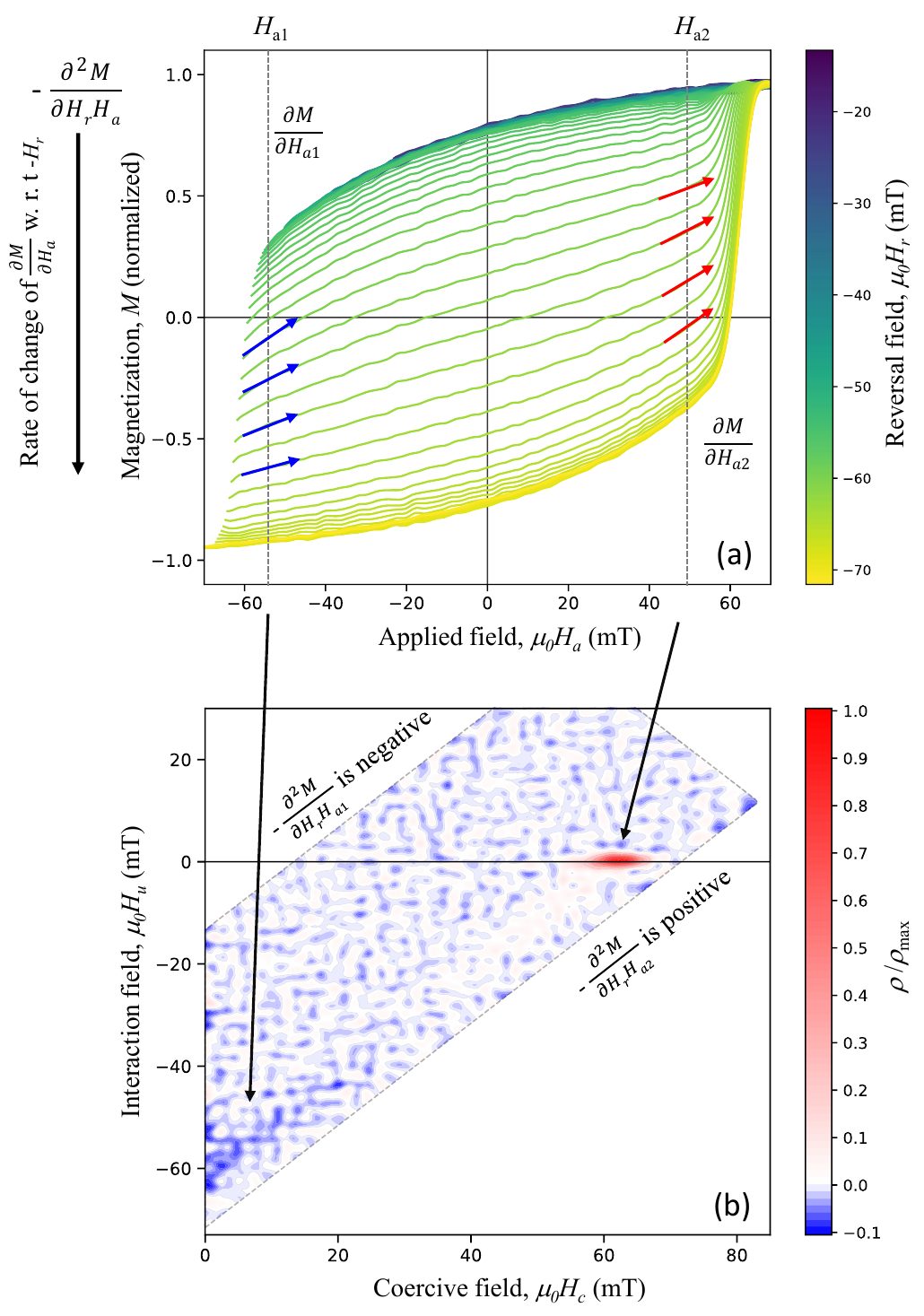"}
\caption{\linespread{1.0} (a) Processed family of FORCs acquired for S1. (b) Corresponding FORC diagram of S1.}
\label{Fig:2}
\end{figure}

\begin{figure}[h!]
\centering
\includegraphics[width=0.9\linewidth]{"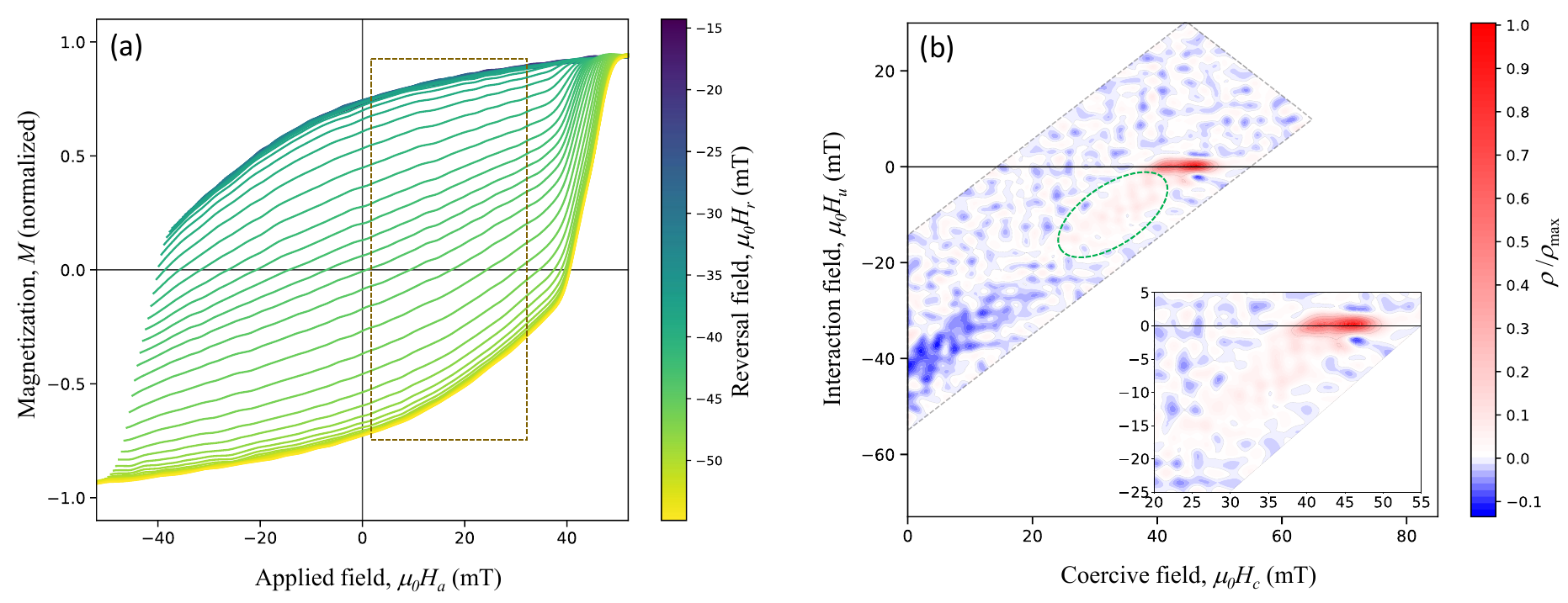"}
\caption{\linespread{1.0} \footnotesize (a) Representative FORCs of S2. (b) Corresponding FORC diagram in the $H_u$--$H_c$ coordinate system. Inset: zoomed-in view of the central peak with a boomerang-shaped feature.}
\label{Fig:3}
\end{figure}

A first-order reversal curve (FORC) diagram is constructed from a series of partial hysteresis loops, commonly referred to as FORCs, as shown in Figure~\ref{Fig:2}(a) and Figure~\ref{Fig:3}(a). The measurement procedure begins by saturating the magnetic system under a sufficiently large positive applied field. The field is then reduced to a selected reversal field, $H_r$. From this reversal point, the applied field is increased again toward positive saturation, and the resulting magnetization trace defines an individual FORC. The magnetization measured at an applied field $H_a$ along a FORC associated with reversal field $H_r$ is denoted $M(H_a, H_r)$, with the constraint $H_a > H_r$. The FORC distribution is obtained from the mixed second-order derivative of the magnetization with respect to the applied field $H_a$ and reversal field $H_r$ and is given by \cite{pike1999characterizing,roberts2014understanding}
\begin{equation}
\rho(H_a, H_r) = -\frac{1}{2} \frac{\partial^2 M(H_a, H_r)}{\partial H_a \, \partial H_r}.
\end{equation}

For physical interpretation, the FORC distribution is commonly displayed as a contour plot in a rotated coordinate system defined by the local interaction field $H_u$ and the coercive or switching field $H_c$. These coordinates are related to the applied and reversal fields through the transformations $H_u = \frac{H_a + H_r}{2}$ and $H_c = \frac{H_a - H_r}{2}$. Because $H_a > H_r$, it follows that $H_c > 0$; therefore, the FORC distribution lies entirely in the right half-plane.

Figure~\ref{Fig:2}(a) shows the processed FORC curves of S1 (AR $= 4.3$, $a = 700$ nm), where the colored minor loops correspond to distinct reversal fields, as indicated by the $H_r$ color bar. The corresponding FORC diagram plotted in the $H_u$--$H_c$ coordinate system is shown in Figure~\ref{Fig:2}(b); the representation in the $H_r$--$H_a$ coordinate system is provided in Figure~S1 of the Supporting Information. Two main features are observed: (i) a negative region in the lower-left part of the diagram and (ii) a pronounced central positive peak. The negative region originates from portions of the FORCs measured at negative applied fields ($H < 0$). This feature arises from a reduction in the magnetization slope, $\partial M / \partial H$, with decreasing reversal field $H_r$ at negative applied fields. The prominent central positive peak is associated with an increase in $\partial M / \partial H$ as $H_r$ decreases at positive applied fields near the dominant switching (coercive) field. For clarity, the slope variations $\partial M / \partial H_{a1}$ and $\partial M / \partial H_{a2}$ at $H_{a1}$ and $H_{a2}$, respectively, are illustrated in Figure~\ref{Fig:2}(a). At $H_{a1}$, the slope $\partial M / \partial H_{a1}$ decreases as $H_r$ decreases, corresponding to the negative region in the FORC distribution. In contrast, at $H_{a2}$, the slope $\partial M / \partial H_{a2}$ increases as $H_r$ decreases, which is associated with the formation of the central peak. The central peak is centered at $H_u = 0$ and extends along the $H_c$ axis, indicating single-domain switching behavior associated with noninteracting or weakly interacting elements \cite{roberts2014understanding}. This suggests that magnetization reversal in this field range is governed primarily by coherent or quasi-coherent rotation along the applied-field direction. The coercive field of the sample is approximately 62~mT.

When the nanomagnet width is increased while the pitch remains the same as in S1 (S2: AR = 3, $a = 700$ nm), a noticeable change appears in the FORC diagram, as shown in Figure~\ref{Fig:3}(b). Although the central positive peak and the negative region remain, an additional asymmetric boomerang-shaped feature develops around the central peak. The lower-left part of this boomerang feature (green circle in Figure~\ref{Fig:3}(b)) originates from FORCs whose reversal fields lie just below the dominant positive switching field (see the highlighted rectangular region in the FORC curves in Figure~\ref{Fig:3}(a)). In this field range, magnetization reversal occurs irreversibly over a narrow field interval, so small changes in reversal field $H_r$ lead to substantial differences in the subsequent FORC return paths. This strong dependence of the magnetization on reversal field produces an enhanced FORC signal, giving rise to the lower-left portion of the boomerang feature \cite{roberts2014understanding,muxworthy2004influence}. The presence of this feature reflects interaction-driven switching processes within the system \cite{roberts2014understanding,muxworthy2004influence}. The coercive field of this sample is reduced to $\sim 45$~mT.

\begin{figure}[h!]
\centering
\includegraphics[width=0.9\linewidth]{"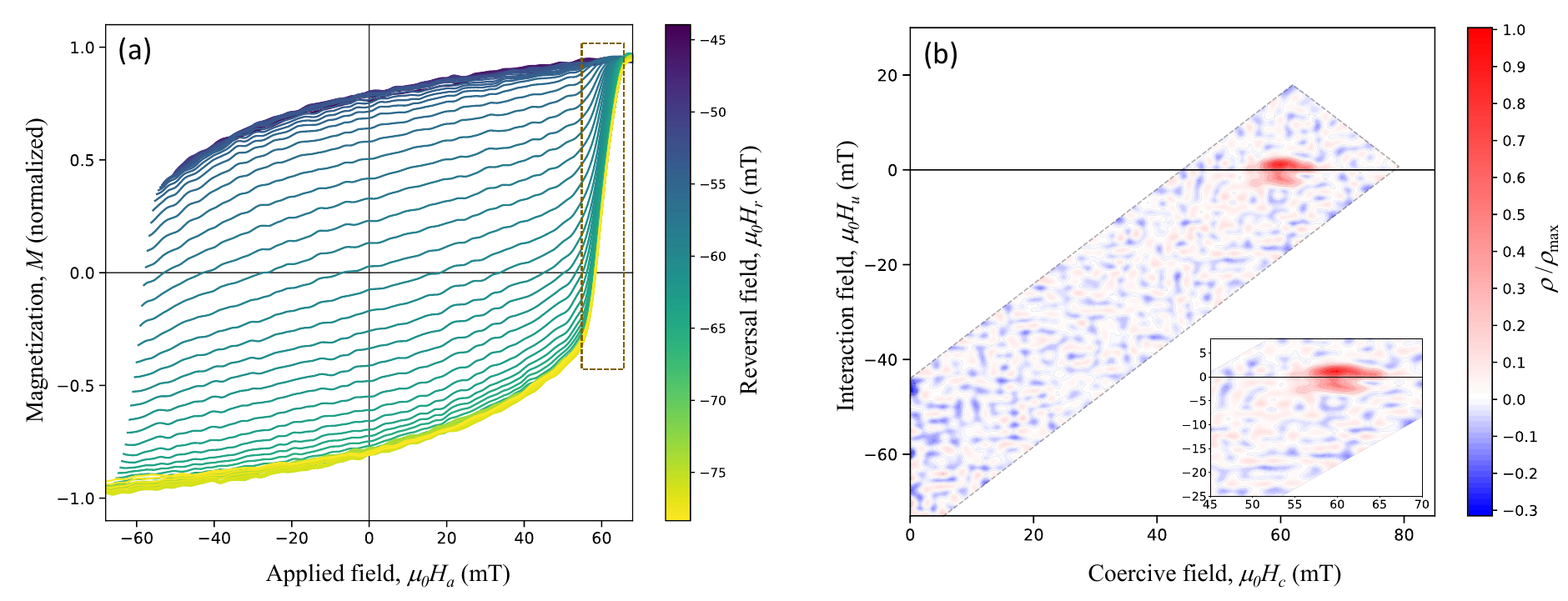"}
\caption{\linespread{1.0} \footnotesize (a) A family of FORCs for S3 with the applied field along the [1,1] direction of the lattice. (b) Corresponding FORC diagram plotted in the $H_c$--$H_u$ coordinate system. Inset: zoomed-in view of the central peak highlighting its vertical spreading.}
\label{Fig:4}
\end{figure}

When the lattice pitch is reduced while the nanomagnet width is kept identical to that of sample S1, a significant change is observed in the FORC distribution of S3. In particular, the central peak becomes elongated along the vertical $H_u$ axis, in contrast to S1, where it extends primarily along the $H_c$ axis (Figure~\ref{Fig:4}(b)). The vertical spreading of the central peak arises from the higher density of FORC curves near the switching field, as indicated by the rectangle in Figure~\ref{Fig:4}(a). This elongation indicates a broad distribution of effective bias (interaction) fields, suggesting that switching events occur over a wide range of local interaction environments. This behavior provides clear evidence of enhanced dipolar interactions in the system, leading to magnetization reversal that is increasingly dominated by inter-island coupling. Despite this, the coercive field of S3 ($\sim 60$~mT) remains very similar to that of S1, indicating that it is governed primarily by intrinsic properties of the individual nano-islands, such as geometry and material characteristics.

\begin{figure}[h!]
\centering
\includegraphics[width=1\linewidth]{"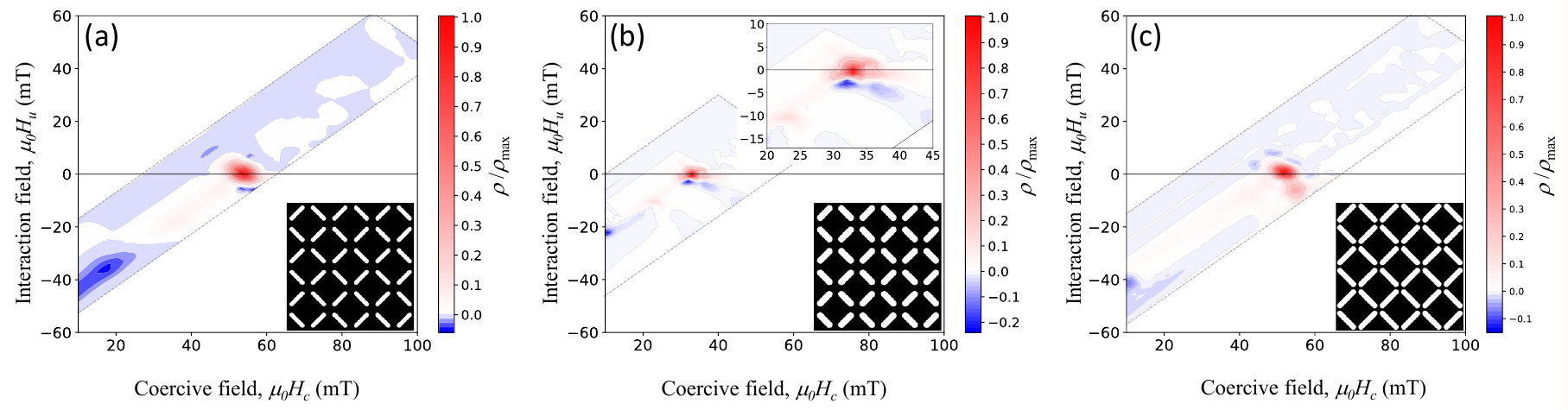"}
\caption{\linespread{1.0} \footnotesize Simulated FORC distributions of Case~I (S1) (a), Case~II (S2) (b), and Case~III (S3) (c). Inset: square array of 36 nanomagnets used in the simulations.}
\label{Fig:5}
\end{figure}

To gain deeper insight into the influence of demagnetization energy on the FORC distributions and the resulting magnetization textures, micromagnetic simulations were performed using MuMax3. A square lattice consisting of 36 nanomagnets was considered (see the inset of Figure~\ref{Fig:5}). In analogy with the experimental conditions, the nanomagnet width and lattice pitch were varied while the nanomagnet length was kept constant.

Three different square-lattice configurations were examined, corresponding directly to the experimental samples: (i) Case~I (S1): $w = 100$~nm, $L = 450$~nm, $a = 700$~nm, and aspect ratio $AR = 4.5$; (ii) Case~II (S2): $w = 150$~nm, $L = 450$~nm, $a = 700$~nm, and $AR = 3$; and (iii) Case~III (S3): $w = 100$~nm, $L = 450$~nm, $a = 550$~nm, and $AR = 4.5$. The nanomagnet dimensions used in the simulations differ slightly from those realized experimentally, primarily because of routine fabrication tolerances inherent to electron-beam lithography. Nevertheless, the qualitative agreement between the simulated and experimental results confirms that the essential magnetic phenomena are insensitive to these minor dimensional variations.

Initially, a random magnetization configuration was assigned to all lattices, followed by energy minimization to obtain the ground state. At equilibrium, the demagnetization energy density for each lattice was calculated and is summarized in Table~\ref{tab:1}. Increasing the nanomagnet width from Case~I (S1) to Case~II (S2) resulted in a modest $\sim 3\%$ increase in the demagnetization energy density ($E_{demag}$). In contrast, reducing the lattice pitch from Case~I (S1) to Case~III (S3) led to a much larger increase of $\sim 40\%$.

\begin{table}[h]
\centering
\begin{tabular}{l|c|c}
\noalign{\hrule height 1.2pt}
 & $E_{demag}$ (J/m$^3$) & Coercive field, $H_c$ (mT)  \\ \hline \hline
Case I (S1) & $2.21\times10^3$ & $\sim$ 55 \\ 
Case II (S2) & $2.28\times10^3$ & $\sim$ 33  \\ 
Case III (S3) & $3.13\times10^3$ & $\sim$ 52  \\ 
\noalign{\hrule height 1.2pt}
\end{tabular}
\caption{Values of $E_{demag}$ and $H_c$ obtained from micromagnetic simulations for Case~I (S1), Case~II (S2), and Case~III (S3).}
\label{tab:1}
\end{table}

Furthermore, FORC diagrams were simulated to validate the influence of demagnetization energy on the reversal behavior. In contrast to the experimental procedure---in which the system is first driven to saturation and then swept to negative fields before each FORC---each simulated FORC was initialized from pre-saved configurations to improve computational efficiency. To preserve field history, ground-state configurations were generated separately and then used as initial states for the minor-loop simulations. The external magnetic field was applied at a small angle of $2.25^\circ$ with respect to the horizontal ($x$) axis, and the corresponding $x$-components of field and magnetization were used for further analysis. The simulated FORC diagrams closely reproduce the key features observed experimentally. For Case~I (S1), a central peak at $H_u = 0$ is observed and elongated along the $H_c$ axis. In Case~II (S2), characteristic boomerang-like features emerge around the central peak. In contrast, for Case~III (S3), the central peak is predominantly elongated along the $H_u$ axis. These observations clearly demonstrate that enhanced demagnetization energy strongly influences switching behavior, leading to a substantial modification of the interaction-field distribution.

\begin{figure}[h!]
\centering
\includegraphics[width=1\linewidth]{"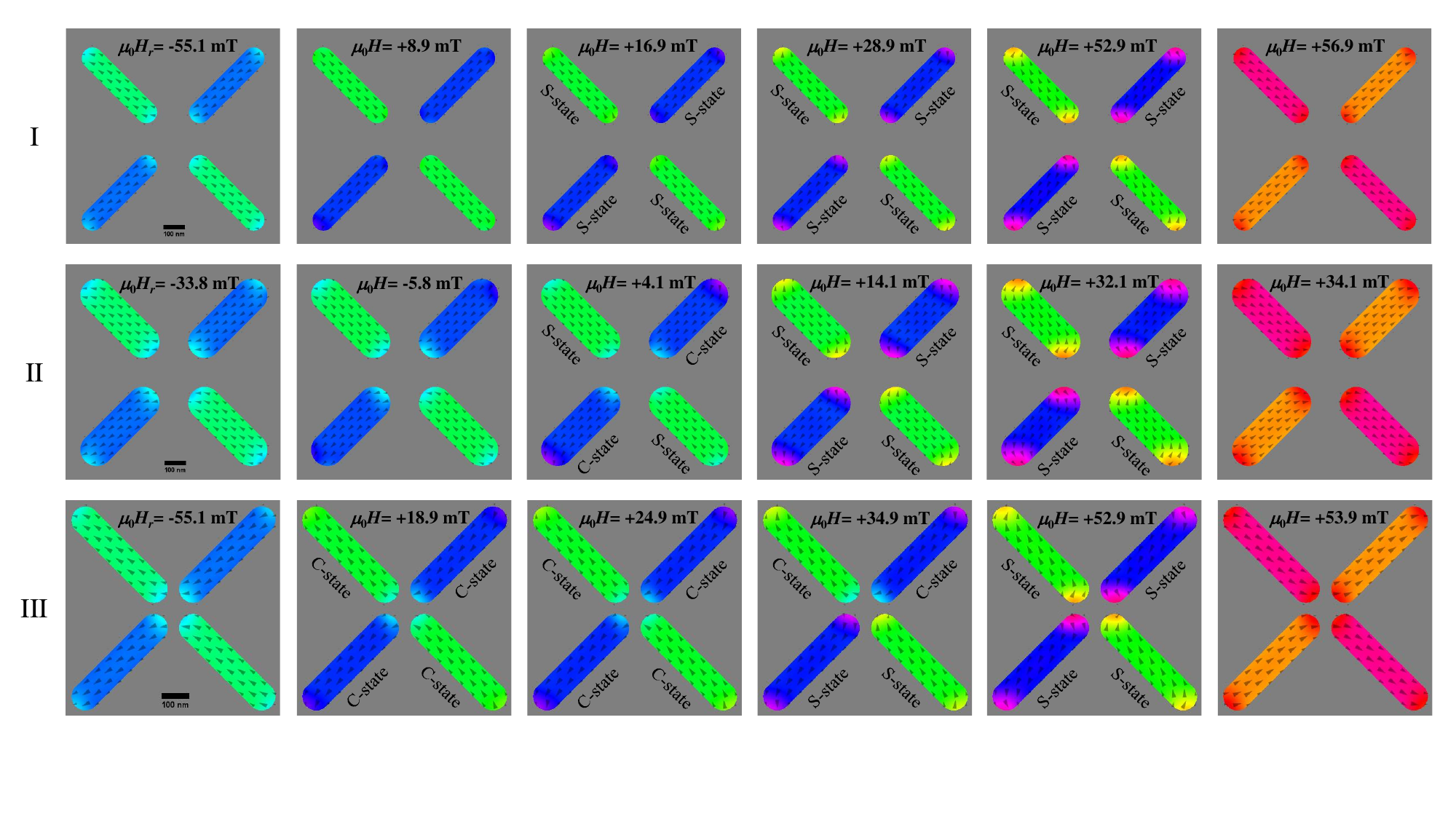"}
\caption{\linespread{1.0} \footnotesize Evolution of magnetic texture in nanomagnets as a function of applied field for Case~I (S1), Case~II (S2), and Case~III (S3). The top, middle, and bottom rows correspond to the simulated counterparts of S1, S2, and S3, respectively. The images show the field-driven reversal sequence from the negative reversal state toward positive saturation, highlighting the distinct intermediate configurations that emerge in each case, including S-domain states, mixed S + C states, and C-state-mediated reversal. The comparison illustrates how changes in geometry and interaction strength modify the reversal pathway and the complexity of the internal magnetic textures.}
\label{Fig:6}
\end{figure}

The internal magnetization textures were also examined for all systems to elucidate the role of demagnetization energy. Figure~\ref{Fig:6} presents the magnetization-reversal behavior of the nanomagnets as a function of applied field for all cases. The systems were initially prepared at a negative reversal field to attain a stable ground state, after which the magnetic field was incrementally applied in the opposite direction until positive saturation was reached.

In Case~I, corresponding to sample S1, the magnetization state at $H_r = -55.1$~mT is considered, with the net magnetization aligned along the $-x$ axis. As the applied field is increased toward positive saturation, the magnetization reverses through an intermediate S-domain configuration. In square ASI, nanomagnets typically behave as Ising-like macrospins because of the strong shape anisotropy of individual islands. In systems with relatively low demagnetization energy, the moments within the nanomagnets rotate in response to the applied field. Initially, moments at the edges begin to rotate, forming an S-domain state (top row of Figure~\ref{Fig:6}, Case~I/S1; see Figure~S2 in the Supporting Information). Once the applied field overcomes the shape anisotropy, the moments switch to the opposite direction. Because inter-island interactions are minimal in this case, all nanomagnets switch coherently with the applied field, giving rise to the symmetric central peak observed in the FORC diagram.

In Case~II, corresponding to sample S2, the system is initialized at $H_r = -33.8$~mT along the $-x$ axis. With increasing applied field, the magnetization reverses through a mixed domain state (S + C-type domains; see Figure~S2 in the Supporting Information), which eventually evolves into purely S-type domains (middle row of Figure~\ref{Fig:6}, Case~II/S2). The slight increase in demagnetization energy enhances interactions among the nanomagnets and stabilizes C-type domain textures alongside S-type domains. As the applied field increases further, the C-type domains transform into S-type domains, resulting in uniform S domains across all nanomagnets. Because the aspect ratio is lower than in Case~I, the shape anisotropy is reduced, which softens the magnetic response and lowers the coercive field (shown in Table~\ref{tab:1}). Consequently, the intra-island effect becomes prominent in this case. This combined effect of softer switching and intermediate C-type domain textures manifests as the boomerang-shaped tail in the FORC distribution.

In Case~III, corresponding to sample S3, the initial magnetization is again taken at $H_r = -55.1$~mT along the $-x$ axis. As the applied field increases, magnetization reverses through a sequence of intermediate C states, mixed (S + C) domains, and finally S-type domains (bottom row of Figure~\ref{Fig:6}, Case~III/S3). Because the aspect ratio is the same as in Case~I, the shape anisotropy and coercive field remain similar. However, stronger dipolar interactions among the nanomagnets induce additional domain textures and intermediate states. The enhancement of these interactions increases both the magnitude and the spatial variability of the local bias fields, producing a broad distribution of effective interaction fields. This appears as a vertical elongation along the $H_u$ axis in the FORC diagram and indicates that magnetization reversal is dominated by inter-island interactions and local configuration-dependent effects, rather than solely by intrinsic switching properties.

\section{Conclusions}
In summary, this study systematically investigates how geometry, lattice pitch, and dipolar interactions govern magnetization reversal in square artificial spin ice systems using FORC measurements and micromagnetic simulations. By varying the nanomagnet width and lattice periodicity, it demonstrates how intrinsic properties and inter-island interactions jointly determine the reversal dynamics. High-aspect-ratio nanomagnets exhibit a symmetric central peak in the FORC distribution, indicative of coherent and relatively uniform switching. In contrast, reducing the aspect ratio lowers the shape anisotropy and moderately increases the demagnetization energy, giving rise to characteristic boomerang-shaped features associated with more complex domain textures. Decreasing the lattice periodicity further strengthens dipolar interactions, leading to a broad distribution of interaction fields. This appears as a vertical elongation of the central peak in the FORC diagram, highlighting the transition from intrinsic- to interaction-dominated switching behavior. Micromagnetic simulations corroborate the experimental findings and reveal the evolution of magnetic textures, providing direct insight into how demagnetization energy and inter-island coupling shape the reversal pathways.

Overall, this study establishes a clear correlation between geometric parameters, interaction strength, and FORC signatures in artificial spin ice systems. These findings highlight FORC analysis as a powerful tool for probing interaction-driven magnetization dynamics and provide practical guidelines for designing ASI architectures with tunable magnetic properties for both fundamental studies and potential applications, including field protocols for neuromorphic computing, fading-memory operation, and reservoir reconfigurability \cite{gartside2022reconfigurable,jensen2024clocked,penty2025controllable,bhandari2025clocking}.

\medskip

\medskip
\textbf{Acknowledgements} \par 
The authors acknowledge financial support from the Carl Trygger Foundation (CTS23:2476). They also thank Myfab Uppsala for providing facilities and experimental support. Myfab is funded by the Swedish Research Council (2020-00207) as a national research infrastructure. The authors are grateful to Dr. Petra E. J\"onsson for valuable discussions. V.K. gratefully acknowledges financial support from the Faculty of Science and Technology at Uppsala University, for a sabbatical stay in 2025 in the group of Prof.\ Peter Schiffer at Princeton University, during which most of this work was carried out, and sincerely thanks Prof.\ Schiffer and his group for their hospitality.

\medskip
\textbf{Conflict of Interest} \par
 There are no conflicts of interest related to this work.

\bibliographystyle{MSP}
\bibliography{References}

\end{document}


\begin{center}

\title{ \textbf{Supplementary Information} \\
Mapping Reversal Pathways and Interaction Fields in Artificial Spin Ice}

\maketitle

\author{Brindaban Ojha $^{1, *}$},
\author{Matías P. Grassi $^{1}$},
\author{Vassilios Kapaklis $^{1, *}$}

\begin{affiliations}
$^{1}$Department of Physics and Astronomy, Uppsala University, Box 516, 75120 Uppsala, Sweden\\
\end{affiliations}

\end{center}

\begin{abstract}

\end{abstract}


\section{FORC diagram in different co-ordinates}

\begin{figure} [H]
	\centering
	\includegraphics[width=0.9\linewidth]{"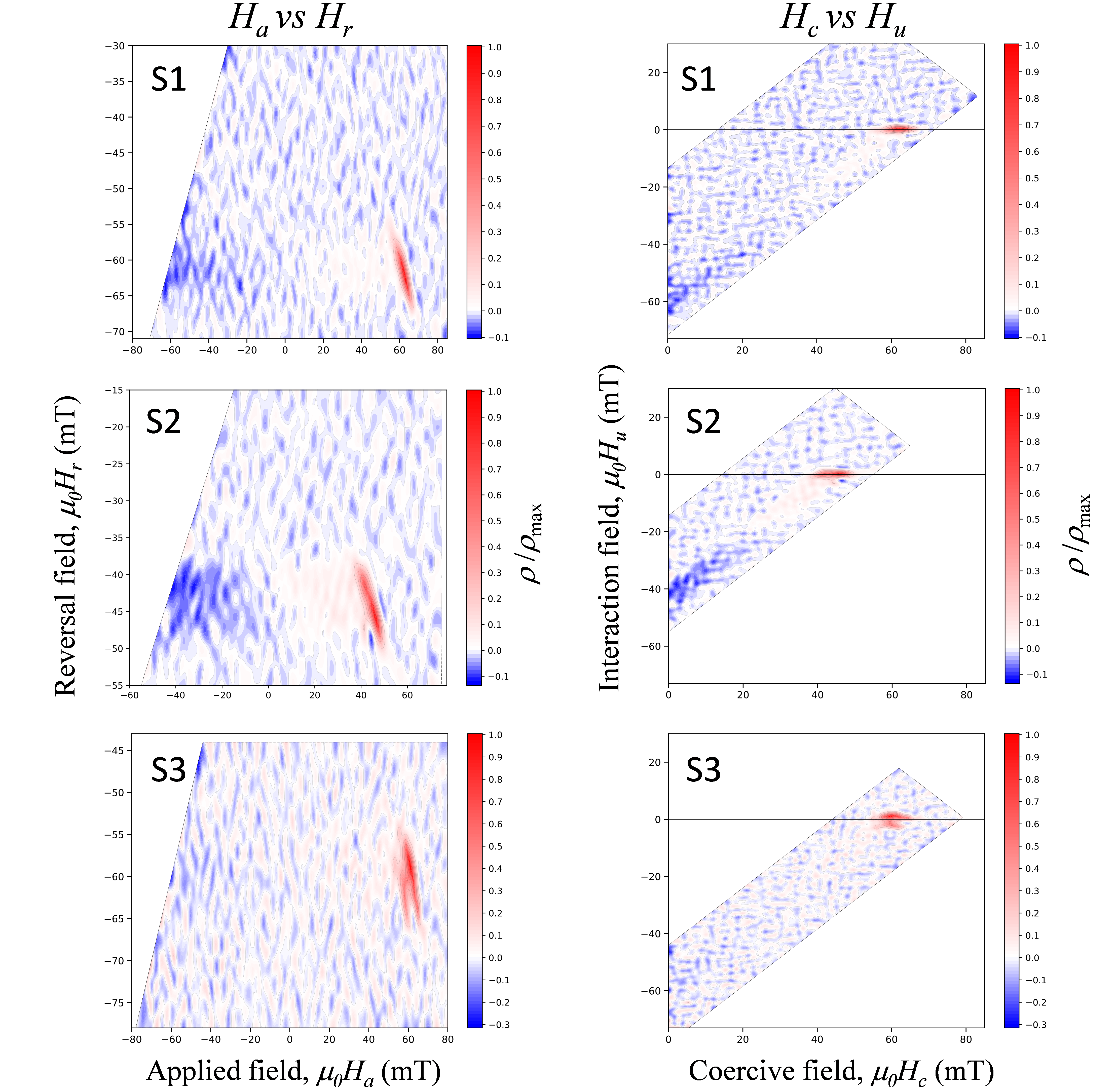"}
    \renewcommand{\thefigure}{S\arabic{figure}}
	\caption{ The FORC diagrams for samples S1, S2, and S3 are presented in both the $H_a$–$H_r$ coordinate system and the rotated coordinate system ($H_c$–$H_u$).}
	\label{fig:Fig_S1}
\end{figure}

\section{Different magnetic domain textures}

\begin{figure} [H]
	\centering
	\includegraphics[width=0.7\linewidth]{"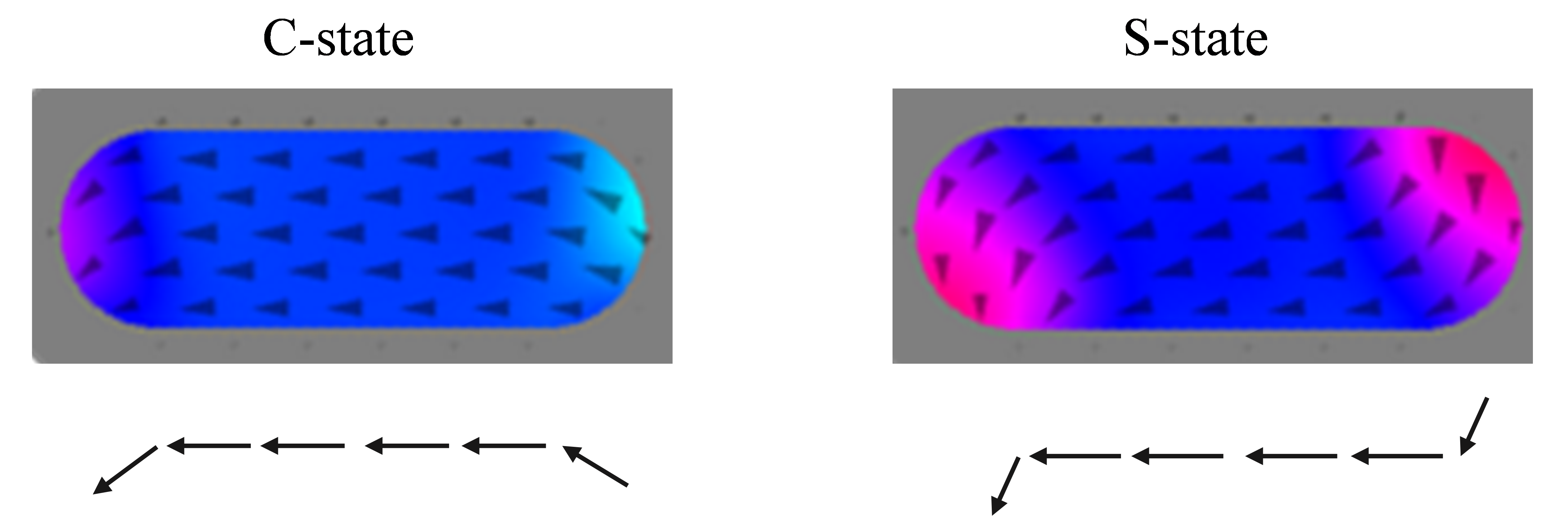"}
    \renewcommand{\thefigure}{S\arabic{figure}}
	\caption{ C- and S-state domain textures arising during the magnetic field sweep. The black arrows indicate the magnetization directions.}
	\label{fig:Fig_S2}
\end{figure}

\bibliographystyle{MSP}